**Origin of threshold current density for asymmetric magnetoresistance in Pt/Py bilayers**

Tian Li[1], Sanghoon Kim[1], Seung-Jae Lee[2], Seo-Won Lee[3], Tomohiro Koyama[4], Daichi Chiba[4], Takahiro Moriyama[1], Kyung-Jin Lee[2,3], Kab-Jin Kim[1, 5†], Teruo Ono[1, 6†]

[1]Institute for Chemical Research, Kyoto University, Uji, Kyoto 611-0011, Japan

[2]KU-KIST Graduate School of Converging Science and Technology, Korea University, Seoul 02841, Republic of Korea

[3]Department of Materials Science and Engineering, Korea University, Seoul 028411, Republic of Korea

[4]Department of Applied Physics, Faculty of Engineering, The University of Tokyo, Bunkyo, Tokyo 113-8656, Japan

[5]Department of Physics, Korea Advanced Institute of Science and Technology, Daejeon 34141, Korea

[6] Center for Spintronics Research Network (CSRN), Graduate School of Engineering Science, Osaka University, Machikaneyama 1-3, Toyonaka, Osaka 560-8531, Japan

†Correspondence to: kabjin@kaist.ac.kr, ono@scl.kyoto-u.ac.jp

An asymmetric magnetoresistance (MR) is investigated in Py/Pt bilayers. The asymmetric MR linearly increases with current density up to a threshold, and increases more rapidly above the threshold. To reveal the origin of threshold behavior, we investigate the magnetic field dependence of the asymmetric MR. It is found that the magnetic field strongly suppresses the asymmetric MR only above the threshold current density. Micromagnetic simulation reveals that the reduction of magnetization due to the spin-torque oscillation can be the origin of the threshold behavior of asymmetric MR.

Thin-film bilayer structures composed of ferromagnetic (FM) and non-magnetic (NM) metals are intriguing systems because of many physical phenomena arising from the spin-orbit coupling such as spin Hall effect (SHE) [1-5], Rashba effect[6-8] and Dzyaloshinskii-Moriya interaction[9-13]. Those spin-orbit-coupling-related phenomena influence spin transport and/or spin dynamics. For instance, when the current flows in a FM/NM bilayer system, the spin Hall effect generates a transverse spin current in NM and induces a spin accumulation at the interface between FM and NM [14]. Since the diffusion of the accumulated spins into FM depends on the relative angle between the magnetization direction of FM and the spin orientation of injected spin current, FM/NM systems can have a unique magnetoresistance (MR), namely the spin Hall MR [15], which is symmetric upon the current polarity flowing in FM/NM systems. Furthermore, it has been recently reported that an asymmetric MR upon the current polarity flowing in FM/NM systems also emerges in FM/NM bilayers, which is caused by either spin accumulation[16] or electron-magnon scattering [17]. Interestingly, the asymmetric MR shows a unique current density dependence: it linearly increases with current density up to a threshold, and increases more rapidly above the threshold[17]. Such a threshold behavior implies that the asymmetric MR has different origin depending on the current density below or above the threshold.

In this letter, we investigate the asymmetric MR under various magnetic fields to reveal the origin of the threshold behavior. We find that the asymmetric MR below the threshold is not affected by the external magnetic field up to 9 T, which is the maximum field that we can apply in our setup. This result is consistent with the scenario that the generation of spin-flip-induced terahertz (THz) magnon, of which energy scale is larger than 1 meV (computed from $g\mu_B B$ with $g = 2$), is responsible for the asymmetric MR below the threshold. On the other hand, the asymmetric MR above the threshold is found to be suppressed by the magnetic field in the examined field range. This implies that much smaller energy scale of magnetization excitation is involved in the asymmetric MR above the threshold. To understand the observed phenomena, we perform the micromagnetic simulation and find that the rapid increase in the asymmetric MR above the threshold can be attributed to the spin-torque-induced magnetization excitation, which lies in the gigahertz (GHz) frequency range. Our results therefore suggest that the asymmetric MR has two different origins, the spin-torque-induced GHz magnon excitation above the threshold and the spin-flip-induced THz magnon excitation independent of the threshold.

Figure 1(a) illustrates the wire structure comprised of Py (NiFe) (4.5 nm)/Pt (5 nm) bilayer. The bilayer films are prepared by the magnetron sputtering and then are patterned into wires by electron beam lithography and Ar ion milling. The width and length of the wire are 300 nm and 10 μm, respectively. An electric current is injected along + $x$ direction, which we define as a positive current. Because of the spin Hall effect in Pt[18], an in-plane charge current generates an out-of-plane spin current, resulting in the spin injection into the Py layer. The red and blue arrows in Fig. 1(a) represent the spin of conduction electrons, whose trajectories are deflected by the spin Hall effect. Since the Pt has a positive spin Hall angle, a positive in-plane current injects conduction electrons of which spin orientation is along the − $y$ direction into the Py.

The longitudinal resistance is measured as a function of the magnetic field $B$ applied along $y$-direction. Figure 1(b) shows the normalized resistance $\Delta R/R_{max}$ as a function of $B$ for various current densities. Here, $R_{max}$ denotes the maximum value of resistance during the $B$ scan and $\Delta R(B) = R(B) - R_{max}$. For a small current (blue line), $\Delta R/R_{max}$ exhibits almost symmetric behavior with respect to $B = 0$, which is observed in typical anisotropic magnetoresistance (AMR)[19] or spin Hall magnetoresistance (SMR) curves[15]. For a larger current, however, the symmetry is broken and the asymmetric MR appears. For $|B| > 20$mT, the $\Delta R/R_{max}$ increases rapidly with the current density $J$ only for a negative $B$, in which the injected spins are antiparallel to the spins of local magnetic moment of FM. This result implies that the interaction between the conduction electron spin and the localized magnetic moment in FM plays a crucial role in the asymmetric MR.

To be more quantitative, we summarize the asymmetric MR, $MR_{asym} \equiv [\Delta R(B = +150$ mT$) - \Delta R(B = -150$ mT$)]/R_{max}$, as a function of current density in Fig. 1(c). The asymmetric MR is found to strongly depend on the current density. It increases linearly with current density up to a threshold ($J_{th} \sim 0.8 \times 10^{12}$ A/m$^2$), and increases more rapidly above the threshold.

To understand the threshold behavior, we investigate the magnetic field dependence of the threshold current density observed in the asymmetric MR. Measurement is performed at $T = 10$ K to suppress the thermal effects. Figure 2(a) shows the observed asymmetric MR as a function of the current density for several external magnetic fields. Three distinct trends are observed. First, the threshold current density $J_{th}$ increases with increasing the external magnetic field. This means that the asymmetric MR above the threshold is suppressed by applying magnetic field. Second, the asymmetric MR below the threshold does not change in the field range investigated. Third, the linear asymmetric MR survives even above the threshold current density. In other words, there is a constant slope in the asymmetric MR for all field values. To see clearly, we plot the threshold current density, $J_{th}$, and the linear slope, $m$, in the asymmetric MR as a function of magnetic field in Fig. 2(b) and 2(c), respectively. It is clear that the $J_{th}$ strongly depends on the magnetic field, whereas the $m$ is insensitive to the magnetic field. The distinct magnetic field dependence for the current density below and above the threshold suggests that the asymmetric MR is governed by two different mechanisms.

It was recently reported that the asymmetric MR emerges in FM/NM bilayers originates from the electron-magnon scattering[17]. In this scenario, spins induced by the spin Hall effect of NM are injected into the FM layer and generate magnons in FM layer via spin-flip process. The resulting magnons have high energies in THz frequency range, and thus induce a resistance change via the electron-magnon scattering. Such a mechanism is consistent with our finding that the asymmetric MR below the threshold is not affected by the magnetic field up to 9T, because THz magnons have an energy of a few meV which is much larger than the Zeeman energy induced by the magnetic field of 9T (~1 meV computed from $g\mu_B B$ with $g = 2$). On the other hand, the asymmetric MR above the threshold is strongly affected by the external magnetic field, implying that another mechanism with a much lower energy scale should be involved. One possible mechanism is the spin torque effect, which generates a magnetization excitation in GHz frequency range[20-22].

To confirm this, we perform the micromagnetic simulation by numerically solving the Landau-Lifshitz-Gilbert equation including the spin torque effect[17,20]. We use the following parameters: The width and thickness of wire are 100 nm and 5 nm, the saturation magnetization $M_S$ is 800 kA/m, the exchange stiffness constant $A_{ex}$ is $1.3\times10^{-11}$ J/m, the spin Hall angle $\theta_{SH}$ is 0.07, the Gilbert damping constant $\alpha$ is 0.015, the temperature $T$ is 10 K, and the unit cell size is $4\times4\times5$ nm$^3$. Figure 3 shows the computed asymmetric magnon number $\Delta N_m(J)$ ($\equiv N_m(+J) - N_m(-J)$), where $N_m$ ($= <M_S-M_y>/M_S$) is proportional to the difference between the saturation magnetization $M_S$ and the $y$ component of the magnetization ($M_y$). That is, $\Delta N_m(J)$ corresponds to the reduction of net magnetization, which results in a resistance change based on the AMR. It is clear that the simulation reproduces the threshold behavior. The simulation result is consistent with the well-known spin-torque-driven magnetic excitation; the $\Delta N_m(J)$ starts to increase rapidly at a current density where the net effective damping becomes zero due to the spin-torque-induced anti-damping effect[23]. This means that the spin-torque-induced reduction of net magnetization and the resulting resistance change is a possible origin for the rapid increase in the asymmetric MR above the threshold. To further confirm the magnetic field dependence of asymmetric MR, we repeat the micromagnetic simulation for various magnetic fields (see Fig. 3). The simulation well reproduces the experimental result: the threshold current density increases with increasing external magnetic field, suggesting that the asymmetric MR above the threshold is induced by the reduction of magnetization due to the spin-torque effect. Note that the simulation cannot reproduce the asymmetric MR below the threshold, suggesting that the asymmetric MR below the threshold cannot be explained by the low frequency (GHz) magnetic excitation induced by the spin torque effect.

To conclude, we investigated the magnetic field dependence of asymmetric MR and found that the field dependence strongly depends on the current density whether it is below or above the threshold current density. Below the threshold, the asymmetric MR was not affected by the external magnetic field up to 9T, which is consistent with the scenario that the high energy THz magnon is responsible for the asymmetric MR. However, the rapid increase in the asymmetric MR above the threshold was strongly suppressed by the magnetic field. This suggests that magnetic excitations with a much lower energy scale is involved in the asymmetric MR above the threshold. Micromagnetic simulation revealed that the spin-torque-induced magnetization excitation is the origin for the rapid increase in the asymmetric MR above the threshold. Our result therefore suggests that the asymmetric MR has two origins: the reduction of magnetization due to the GHz magnon generation by the spin-torque effect and the electron-magnon scattering due to the THz magnon excitation by spin-flip process.

This work was partly supported by JSPS KAKENHI Grant Numbers 15H05702, 26870300, 26870304, 26103002, 25220604, 2604316 Collaborative Research Program of the Institute for Chemical Research, Kyoto University, the Cooperative Research

Project Program of the Research Institute of Electrical Communication, Tohoku University, and R & D project for ICT Key Technology of MEXT from the Japan Society for the Promotion of Science (JSPS). KJK was supported by the National Research Foundation of Korea (NRF) grant funded by the Korea government (MSIP) (No. 2017R1C1B2009686). KJL was supported by the National Research Foundation of Korea (NRF-2015M3D1A1070465, NRF-2017R1A2B2006119).

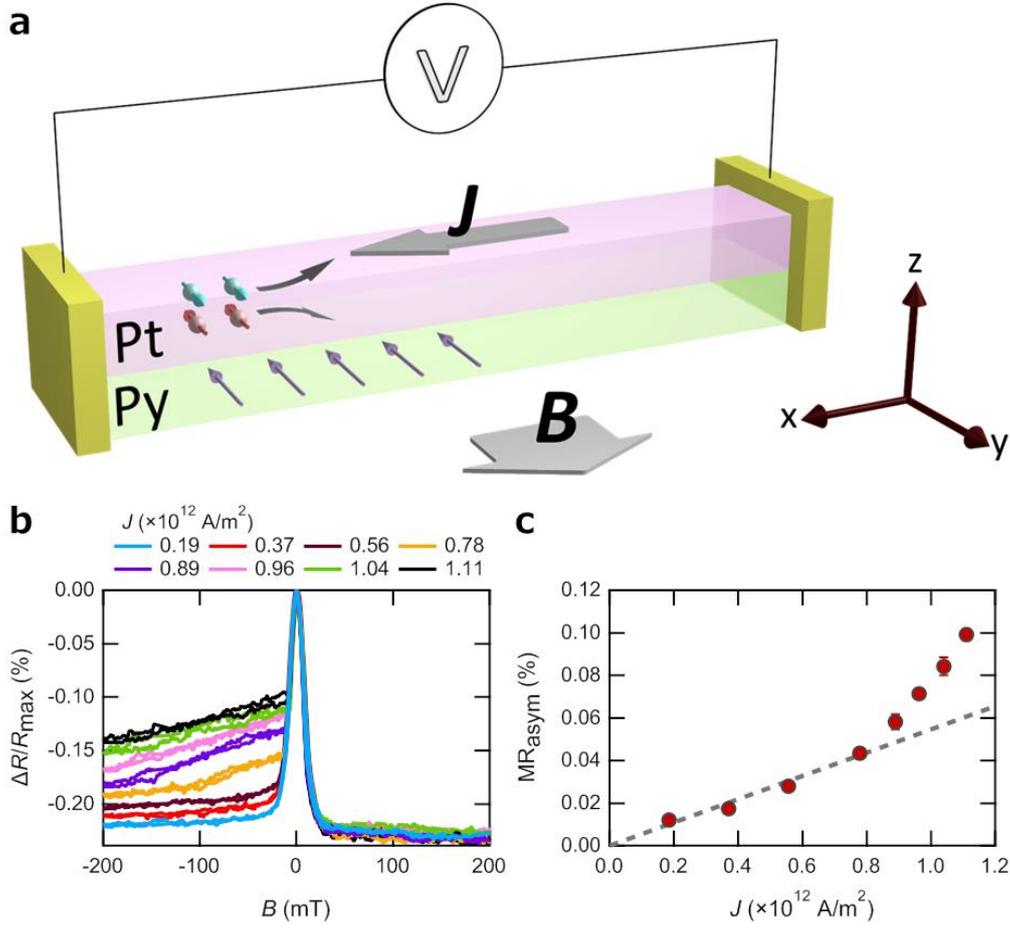

**Figure 1.** (a) Schematic illustration of the Py(5 nm)/Pt(5 nm) bilayer wire with the definition of external magnetic field ($B$) and current density ($J$). The blue and red arrows in Pt layer represent the spin of conduction electrons. The arrow in Py layer represents the spin of local magnetic moments. Note that the spin direction is the opposite to the direction of magnetic moment because of the negative electron charge. (b) Normalized resistance $\Delta R/R_{max}$ as a function of magnetic field for various current densities. (c) Asymmetric magnetoresistance, with the definition of $MR_{asym} \equiv [\Delta R(B = +150 \text{ mT}) - \Delta R(B = -150 \text{ mT})]/R_{max}$, as a function of current density.

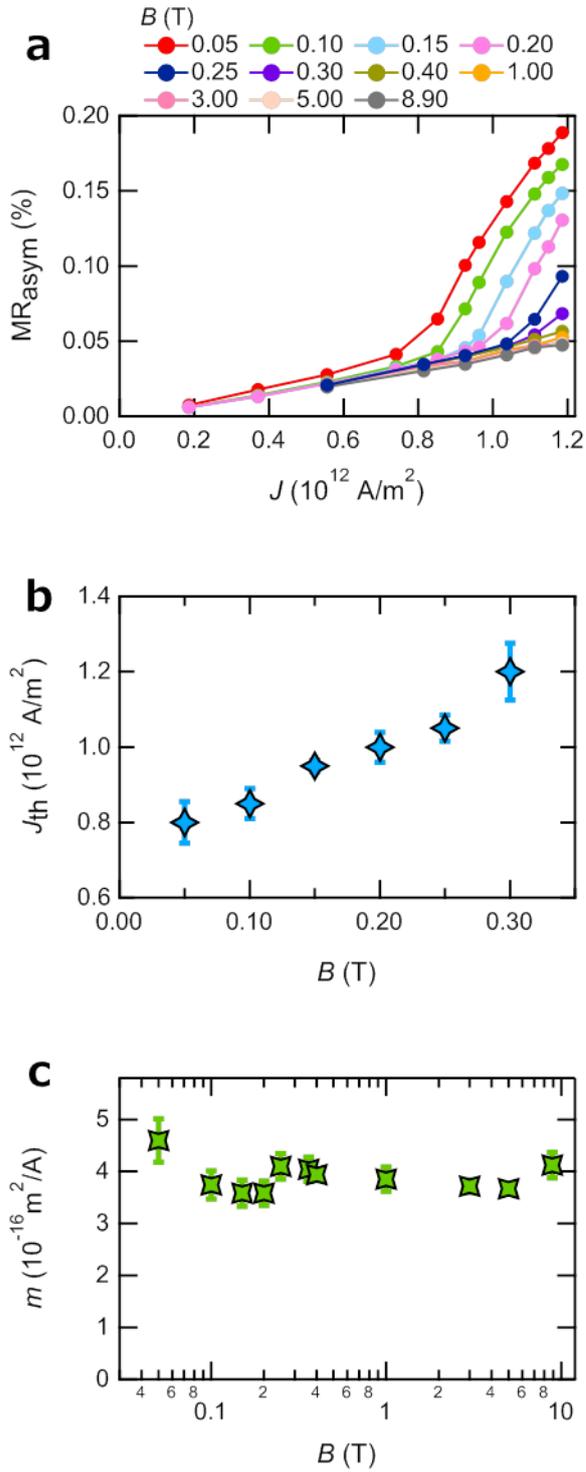

**Figure 2** (a) Asymmetric magnetoresistance, MR$_{asym}$, with respect to the current density for several magnetic fields. (b) Threshold current density, $J_{th}$, as a function of magnetic field. (c) Linear slop, $m$, as a function of magnetic field.

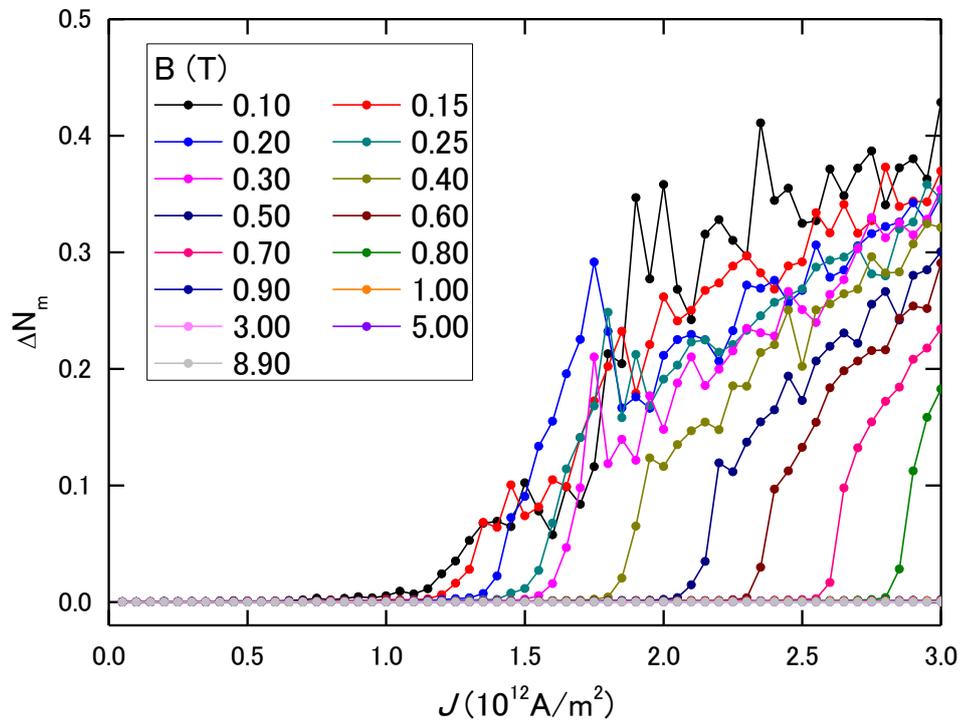

**Figure 3** Computed magnetic field dependence of asymmetric magnon number $\Delta N_m$ as a function of $J$.